# A simple evaluation of global and diffuse Luminous Efficacy for all sky conditions in tropical and humid climate

A. H. Fakra, H. Boyer, F. Miranville, D. Bigot
Building Physics and Systems Laboratory (LPBS)
University of La Reunion
fakra@univ-reunion.fr

**Abstract**
This paper presents initial values of global and diffuse luminous efficacy at Saint-Pierre (Reunion Island). Firstly, data used were measured for a period of 6 months, from February to June 2008. During this period, all defined day-types have been studied (clear, cloudy and intermediate). Generally, the meteorological database of Reunion does not contain information for illuminance values. On the other hand, the local meteorological center has a 60-years-old database for solar irradiance (W.m$^{-2}$). So it is important to determine Luminous Efficacy in order to find illuminance from solar irradiance (or luminance from solar radiance). The measured data were analyzed, and empirical constant models were developed and presented in this paper in order to determine luminous efficacy under different sky conditions. A comparison between these empirical constants (models) and existing models has been made. The method used to define sky conditions (overcast, intermediate and clear) and day-types characterization as well as classification will be presented in this work.

## 1. Introduction

Daylighting is recognized as an important and useful strategy in the design of energy efficient buildings. It is particularly useful in hot climates since it reduces the use of artificial lighting and thus electricity demand. Luminous efficacy is the parameter used to determine outdoor illuminance. In fact, many regions of the world work on this parameter to meet the needs imposed by problematic environments (from a meteorological point of view). For example, illuminance values (in lux or lm.m$^{-2}$) can be simulated from the solar irradiance (W.m$^{-2}$) and luminous efficacy. Therefore, it is important to know theses values of luminous efficacy for a given location and this is the reason for this study at Reunion Island.

Most meteorological stations in La Reunion have a sizeable database for solar irradiance (going back about 60 years) but these stations unfortunately do not have illuminance data.

Another reason that prompts us to study these luminous efficacy values is that simulation software for buildings needs specific climatic data, particularly on outdoor illuminance.

Hence, luminous efficacy models introduced in the software enables illuminance values simulation using solar irradiance databases.

Firstly, empirical (or experimental) constant luminous efficacy (global and diffuse models) for this location (at 21°19'S, 55°28'E) is introduced. Materials, methods and experimental procedures are then presented. Results obtained confirm the accuracy of models.

This work was done for a PhD thesis [1] and results obtained are used in CODYRUN [2] simulation software for predicted illuminance values in Reunion Island.

## 2. Literature review

The luminous efficacy of daylight (in lm/W) is defined as the ratio of daylight illuminance to solar irradiance. With the availability of the measured horizontal and/or vertical solar data, luminous efficacy for horizontal and/or vertical surfaces can be determined respectively. This gives building designers an indication of the relationship between daylight illuminance and solar irradiance. The bibliography is restricted to horizontal luminous efficacy cases only.

This evaluation method of luminous efficacy is an experimental one: it is the ratio between outdoor illuminance and solar irradiance data (or luminance/radiance ratio).

Various research teams around the world had carried out similar studies to determine the luminous efficacy constants.

Among the earliest are the works of Pleijel [3] in Scandinavia, Blackwell [4] at Kew, England, Dogniaux [5] in Uccle, Belgium and Drummond [6] in Pretoria, South Africa. Blackwell concluded that global efficacy varied between 105 and 128 lm/W for average sky conditions. Although no diffuse efficacy values were available at that time, hence a combination of Backwell's irradiance measurements with corresponding illuminance values for Kew (outlined by McDermott and Gordon – Smith for nearby Teddington), [7] led Hopkinson [8] to propose a constant value of 125 lm/W for diffuse efficacy. Muneer and Angus (respectively [9] and [10]) suggested all – sky, average global and diffuse luminous efficacy values of 110 and 120 lm/W, respectively. Darula and Kittler [11], Joseph [12] or Dumortier [13] has given more information on the comparison of the different luminous efficacy found all over the world.

According to Darula [14], each region has its own constant of luminous efficacy. Indeed, in his work, the author demonstrated that these variables are heavily dependent on local climate (precipitation, temperature, etc.), geography location (latitude, longitude, etc.) and meteorological phenomena observed on the site (hurricanes, volcanoes, etc.).

Generally, the maximum average value for global luminous efficacy is 180 lm/W (in Norway) and the minimum average value is 98 lm/W (in Japan) [13]. More recent research in this field has been carried out for Cucumo [15] for different locations and Ana [16], Spain.

In this paper, similar work has been done to determine the constant values for Reunion Island.

### 3. Methodology and experimental overview

The experimental station is located on the rooftop of an experimental test cell located at the University of Technology of Saint Pierre (University of Reunion) and is free from obstructions. In 2008, an experimental station was set up to measure horizontal solar irradiance and outdoor illuminance. The instrumentation of the station consists of two pyranometers and two illuminance sensors for horizontal solar irradiance and daylight illuminance data measurements. All sensors are installed on the rooftop in a position which is relatively free from any external obstructions, and is readily accessible for inspection and general cleaning.

#### 3.1 Location of the meteorological station

The experimental station is located at 21°19'S, 55°28'E, and 70,40m above sea level (see Fig. 1 and 2). The field environment consists of an experimental set up (the LGI cell [17]) and meteorological station (see Fig. 3). Illuminance sensors fitted to the roof of the LGI cell. Measurements of solar irradiance were given by the pyranometer of the local meteorological station.

**Fig. 1. Global view of the experimental platform**   **Fig. 2. Photo of the experimental platform**

#### 3.2 Experimentation set-up

Data collection starts before sunrise and ends after sunset. All measurements are referred to using true solar time (TST). Pyranometers (CM11), manufactured by Kipp & Zonen, was used to carry out the measurements of solar irradiance on a horizontal plane. The diffuse irradiance pyranometer was fitted with a shadow ring 55 mm wide and 620 mm in diameter, which shades the thermopile from the direct beam. The shadow ring is painted black to minimize the effect of multiple reflections. The two pyranometers were connected to a data acquisition system, which calculated the irradiance at intervals of one minute. The data from the system were sent by cable to a microcomputer for storage. The measurements of global and diffuse outdoor illuminance on a horizontal plane were made by means of two illuminance sensors (see Fig. 3). The two silicon photovoltaic cells have cosine and colour correction. Again, a shadow ring was used to stop the direct component to obtain the diffuse illuminance measurement. The measured results were converted into digital signals by an analogue to digital converter before being transferred into a microcomputer for storage. These data were measured at one-second intervals and averaged over one-minute intervals. For measurements of overall illuminance (i.e., direct beam, sky – diffuse and ground - reflected components), one illuminance sensor of the same model type was used. The sensor is connected to another analogue to digital converter and linked to a computer for data conversion and data logging. The data collection method was the same one as that for horizontal illuminance measurements. Quality control tests were carried out to eliminate spurious data and inaccurate measurements resulting from the cosine response error of the pyranometers and illuminance sensors.

Irradiance was measured using a pyranometer (CM11 Kipp & Zonen secondary standard pyranometer) connected to a data logger of the CAMPBELL brand; while measurements associated with illuminance (FLA613VLM luxmeter and corresponding data logger) were done using equipment by the AHLBORN Company. Synchronization of the two experiments was necessary in order to get results of measurements for illuminance and irradiance simultaneously. Their respective manufacturers provide technical specifications for each instrument. Table. 1 And 2 show these errors values.

| Spectral range | 305-2800 nm (50% points) |
|---|---|
| Sensitivity | 4-6 $\mu$V/W/m$^2$ |
| Inpedance (nominal) | 700-1500 $\Omega$ |
| Response time (95%) | 15 sec. |
| Non-linearity | $< \pm 0.6$ % ($<1000$ W/m$^2$) |
| Temp. dependence of sensitivity | $< \pm 1$ % (-10 to +40 °C) |
| Directional error | $< \pm 10$ W/m$^2$ (beam 1000 W/m$^2$) |
| Tilt error | None |
| Zero-offset due to temp. changes | $< \pm 2$ W/m$^2$ at 5 K/h temp. change |
| Operating temperature | -40°C to +80°C |
| ISO-9060 Class | Secondary Standard |

**Tab. 1. CM11 pyranometer specifications.**

| Measuring range | 0 – 170 Klux |
|---|---|
| Spectral sensitivity | 360 – 760 nm |
| Dome | PMMA |
| Cosine error | $< 3$ % |
| Linearity | $< 1$ % |
| Absolute error | $< 10$ % |
| Directional error | $< \pm 10$ W/m$^2$ (beam 1000 W/m$^2$) |
| Operating temperature | -20°C to +60°C |

**Tab. 2. FLA613VLM luxmeter specifications**

**Fig. 3. Global illuminance sensor**

### 3.3. Methods and experimental data

Hubbard [18] demonstrated that the length of data series should be more than one year to characterize a seasonal pattern. Gueymard [19] recommended that, in order to validate irradiance-estimation models, long-term dataset is needed. Data collected every minute for 6 months (February – July) on horizontal plane, measured at the city of Saint-Pierre, were used for the analysis. Two seasons (winter and summer) were observed during this period of the year. In fact, in Reunion Island February correspond to summer season and June winter season.

It is often noticed some short periods of missing data due to various reasons including instrumentation malfunctioning and power failure. Considerable effort was made to obtain 12000 and 23000 minute reading, respectively, for each global and diffuse solar irradiance and daylight illuminance. Graphical representation is a simple and direct approach to analyze and interpret the recorded solar data.

### 3.4. Classification of day-type conditions

Sky conditions are generally classified into three categories, namely: clear, intermediate and overcast. It is important to have information about the occurrence of clear and cloudy days. In fact, the originality of this study is that three typical day-types (or day conditions) have been defined instead of three sky conditions: intermediate (Fig. 5), clear (Fig. 6) and overcast (Fig. 7) days. Global illuminance (sum of direct and diffuse illuminance) is the only variable used to classify the day-types defined. Indeed, testing classification models on diffuse illuminance cannot provide information on the day-type. These day-types (clear, intermediate and overcast) were observed during the 6 months measurement.

Classification assessment began at sunrise and finished at sunset. The frequency distribution of each sky type determines the day classification.

**Fig. 4. Global (Gh), diffuse (dh) and direct (Dh) fluctuations (irradiance) for three typical days**

**Fig. 5. Fluctuation of global (Gh) irradiance for an intermediate day**

**Fig. 6. Fluctuation of global irradiance for clear day**

**Fig. 7. Fluctuation of global irradiance for overcast day**

Exploration of the scientific bibliography gerund index (usually used for determine sky type) was used to determine the three typical day-types defined for this work. Then, selection and testing of this index were made from the defined day conditions. To characterize the day conditions, two different clearness indices (the Perez or Kittler model), a brightness index (the Perraudeau model) and a recently simplified index (Sky Ratio) were used.

In 2002, Kittler finally showed that $K_T$ was not appropriate for the study of classification of sky types [20]. However, it is important to determine whether this index could be applied to classify day – types or not. Other indices are generally used for studies of classification sky types. The aim of this study is also to verify the possibility of using the indices for the characterization of sky types on day-types, which were previously defined.

The most important reason for selecting this index is that the meteorological station laboratory can give input (variables) data and parameters to determine the index. Classification of the daylight data into three-day conditions is given by the four indexes shown in Table 3.

- $G_h$: The global horizontal terrestrial-irradiance (W.m$^{-2}$)
- $d_h$: The diffuse horizontal terrestrial-irradiance (W.m$^{-2}$)
- $d_{es}$: The beam normal terrestrial-irradiance (W.m$^{-2}$)
- $Z$: The zenithal angle (rd)
- $G_o$: Extraterrestrial horizontal solar irradiance or solar constant (W.m$^{-2}$)
- $\gamma$: Solar elevation or altitude angle (°)

| Index Name | Symbols | Analytical Form | Ref. |
|------------|---------|-----------------|------|
| *Clearness* | ε | (1) | [21] |
| *Brightness* | $I_N$ | (2) | [22] |
| *Kittler* | $K_T$ | (3) | [23] |
| *Sky Ratio* | SR | (4) | [24] |

**Tab. 3. Index used to characterize day-types (overcast, clear and intermediate).**

Analytical form index and respective typical conditions to characterize day-types is given by:

- Perez's Clearness index ε :

$$\varepsilon = \frac{\left(\dfrac{d_h + d_{es}}{d_h}\right) + 1.041 \times Z^3}{1 + 1.041 \times Z^3} \qquad (1)$$

Tests conditions are:

$$\begin{cases} \varepsilon \leq 1.23 \Rightarrow \text{Overcast} \\ 1.23 < \varepsilon < 4.5 \Rightarrow \text{Intermediate} \\ \varepsilon \geq 4.5 \Rightarrow \text{Clear} \end{cases} \qquad (2)$$

- Perraudeau's Brightness index $I_N$ :

$$I_N = \frac{1 - \left(\dfrac{d_h}{G_h}\right)}{1 - \left(0.12037 \times \dfrac{1}{(\sin\gamma)^{0.82}}\right)} \qquad (3)$$

This index is classified in categories: overcast, intermediate overcast, intermediate mean, intermediate blue and clear. In our study, we have combined the three intermediate sky categories into a single category. This allows us to work with three day-types only, instead of five, and they are:

$$\begin{cases} 0 \leq I_N \leq 0.05 \Rightarrow \text{Overcast} \\ 0.05 < I_N < 0.9 \Rightarrow \text{Intermediate} \\ 0.9 \leq I_N < 1 \Rightarrow \text{Clear} \end{cases} \qquad (4)$$

- Kittler's index $K_T$ :

$$K_T = \frac{G_h}{G_o} \qquad (5)$$

Tests conditions are:

$$\begin{cases} 0 < K_T \le 0.3 \Rightarrow \text{Overcast} \\ 0.3 < K_T \le O.65 \Rightarrow \text{Intermediate} \\ \qquad K_T > 0.65 \Rightarrow \text{Clear} \end{cases} \qquad (6)$$

- Sky Ratio index $SR$ :

$$SR = \frac{d_h}{G_h} \qquad (7)$$

Tests conditions are:

$$\begin{cases} \qquad SR \ge 0.8 \Rightarrow \text{Overcast} \\ 0.3 < SR < 0.8 \Rightarrow \text{Intermediate} \\ \qquad SR \le 0.3 \Rightarrow \text{Clear} \end{cases} \qquad (8)$$

The application of these indices on different day-types (defined previously) and the comparison of the results obtained between the selected sky classifications indices are given in Fig. 8. For each index, associated tests conditions, enables us to define day-types (see tests conditions for relations (2), (4), (6) and (8)). Finally, the reliability of each index was checked, which describes day-types previously defined.

Analysis of the graphs showed that:

- Using Perraudeau's index, it is difficult to differentiate intermediate days from clear days. In fact, the real classification of sky indices proposed by the author was not taken into account in our day-to-day classification hypothesis (day – types defines);

- Perez's index gives acceptable results for overcast (around 80% success) and clear (around 65% success) days but cannot characterize correctly intermediate days. In fact, values obtained for this index fluctuate considerably in the case of intermediate days (between 1 and 11). The value of this index is too high for our classification study;

- Results obtained in the case of SR index are satisfactory. Success rates of 81%, 98% and 60% was observed in the case of overcast, clear and intermediate days respectively;

- Finally, it is noted that Kittler's index does not suit these day-types. The only variable in the analytic form of the index is the global horizontal terrestrial-irradiance. The extraterrestrial solar irradiance stays constant. So, it is not possible to characterize correctly random cloud phenomenon for each defined day (overcast, intermediate and clear).

To conclude, Perez's index can characterize day-types (defined previously) up to a certain extent (only clear and overcast days) in contrast with the SR index that is more appropriate to characterize day-types defined. Kittler's and Perraudeau's indices are not suitable to Reunion Island.

**Fig. 8. Results of comparisons and application of different indices for day-type classification (clear, intermediate and overcast)**

## 4. Comparisons of luminous efficacy models with measurements (Results)

### 4.1. Statistical errors indicators

The accuracy of the constants models was determined using statically indicators: the coefficient of determination $R^2$, the Mean Bias Deviation (MBD) and the Root Means Square Deviation (RMSD). MBD demonstrates the model's tendency to underestimation or overestimation. RMSD offers a deviation measure from the predicted values in the relation to the measured values. Using the correlation coefficient $R^2$, the MBD, and the RMSD determined the accuracy of the models as statistical estimators [25]. They are defined as follows:

$$RMSD = \left(\frac{1}{E_{mean}}\right) \times \left[\frac{\sum\limits_{i=1}^{N}\left(E_{mod,i} - E_{meas,i}\right)^2}{N}\right]^{1/2} \quad (9)$$

and

$$MBD = \frac{\sum\limits_{i=1}^{N}\left(E_{mod,i} - E_{meas,i}\right)}{N \times E_{mean}} \times 100 \quad (10)$$

Where:

$E_{meas,i}$ : Measured value of the dependent variable corresponding to a particular set of values of the independent variables. In this study, it is the measured values for horizontal illuminance / irradiance;

$E_{mod,i}$ : Predicted dependent variable value for the same set of independent variables above (these values are obtained from the model), In this case, it is the illuminance predicted values from the constant (global or diffuse) luminous efficacy;

$E_{mean}$ : Mean value of the dependent variable testing data set and N (number of records of data in the testing set) or the $E_{meas,i}$ mean values;

In order to increase accuracy, some statistical indicators also need to be defined as follows:

$$R^2 = \frac{\sum\limits_{i=1}^{N}\left(E_{mod,i} - E_{mean}\right)^2}{\sum\limits_{i=1}^{N}\left(E_{meas,i} - E_{mean}\right)^2} \quad (11)$$

Where $R^2$ is the coefficient of correlation.

The definition of the following value at Mean Percentage Deviation (MPD) and Root Mean Square percentage deviation (RMS) between the calculated data and the experimental data, is given by [15]:

$$MPD = \frac{\sum\limits_{i=1}^{N}\left[\frac{\left(E_{mod,i} - E_{meas,i}\right)}{E_{meas,i}}\right]}{N} \times 100 \quad (12)$$

and

$$RMS = \sqrt{\frac{\sum\limits_{i=1}^{N}\left[\frac{\left(E_{mod,i} - E_{meas,i}\right)}{E_{meas,i}} \times 100\right]^2}{N}} \quad (13)$$

*4.2. Simplified correlations of luminous efficacy for Saint-Pierre (a simple constant luminous efficacy in Reunion Island)*

Figs. 9 and 10 show a linear relation between illuminance and irradiance, and consequently, the following simplified correlations of empirical average constant diffuse and global luminous efficacy for Saint-Pierre have been obtained by :

$$K_d = \frac{I_d}{d_h} = 139.98 \ (lm/W) \qquad (14)$$

With: MPD= 0.41 %, RMS=6.95 %

And,

$$K_g = \frac{I_h}{G_h} = 121.5 \ (lm/W) \qquad (15)$$

With: MPD= 8.58 %, RMS=13.69 %

Where $K_d$ is the empirical constant diffuse luminous efficacy, $K_g$ is the empirical constant global luminous efficacy, $I_d$ and $I_h$ are respectively the diffuse and global terrestrial-illuminance (in lux) on the horizontal surface. $d_h$ and $G_h$ are respectively the diffuse and global terrestrial-irradiance (in $W.m^{-2}$).

**Fig. 9. Diffuse illuminance as a function of diffuse irradiance on the horizontal surface at Saint-Pierre**

**Fig. 10. Global illuminance as a function of global irradiance on the horizontal surface at Saint-Pierre**

Diffuse luminous efficacy value remains constant during the day, except at sunrise and sunset where it varies considerably. Indeed, this value is function of the solar elevation. This explains the errors obtained when reducing this value to a constant.

The results obtained for global luminous efficacy showed a fairly stable value throughout the day. It is true that this value is related to direct and diffuse luminous efficacy, and that these variables dependent on solar elevation (refer to [26], [27] and [28]) but, similar studies showed that, generally, the diffuse luminous efficacy increases when the direct luminous efficacy decreases (and vice versa) in proportions that depend on the aerosol content of the atmosphere [13]. Therefore, these two (direct and diffuse luminous efficacy) values tend to an equilibrium value (global luminous efficacy). This probably explains the small variations observed in the experimental value of global luminous efficacy (dependent direct and diffuse luminous efficacy). Errors obtained from the constant global luminous efficacy can be explained by the fact that this value varies solely on solar altitude. Indeed, other parameters such as the amount of aerosol contained in the atmosphere or the distribution of clouds in the sky can significantly influence this constant.

By taking experimental errors (of +/- 1-3%) into consideration (given by the material manufacturer), it is concluded that the luminous efficacy values obtained experimentally are reasonable.

In addition, the constants found are in conformity with international references. For example, the value of global luminous efficacy was found to be 121.5 lm /W. This corresponds to a value between 98-180 lm/W (see reference [13]).

**5. Application: simulation in CODYRUN**

CODYRUN is a dynamical building simulation code developed by the L.P.B.S. laboratory [2]. The software uses a multiple-model approach to simulate the building system (thermal, airflow and humidity aspects in multi-zones buildings). The laboratory developed this program for two types of users:

- Building designers and operators

- Building physics researchers

Thereafter, several works have been investigated to improve the computer tools. Thus, in this scope, many physical models were introduced then tested into the software in order to make it even more powerful and more diversified (reference [29], [30] and [31]).

Recently, improvements were made to the calculation code for simulating exterior daylighting from the meteorological database. In fact, two constants (global and diffuse) luminous efficacy to simulate illuminance values were used. Figs. 11 and 12 illustrate comparisons between the simulated results and measured ones obtained at Saint-Pierre, in the case of comparison for diffuse illuminance (see Fig. 11), showing four successive random days. In the case of comparison for global illuminance (see Fig. 12), three typical days were selected (clear, intermediate and overcast) to compare measured and simulated values.

**Fig. 11. Comparisons between measured and CODYRUN-simulated constant diffuse illuminance values at Saint-Pierre**.

**Fig. 12. Comparisons between measured and CODYRUN-simulated constant global illuminance values at Saint-Pierre**.

Errors between measurements and simulation are:

- RMSD = 5.04 %, MBD= 0.16 % and $R^2$= 0.75 for the diffuse illuminance comparison and,
- RMSD = 5.30 %, MBD= -1.45 % and $R^2$= 0.74 for the global illuminance comparison.

These results showed that the constant model of global and diffuse luminous efficacy predicted better illuminance values at Saint-Pierre.

Outdoor daylighting obtained from luminous efficacy allows the simulation of indoor daylighting values at any point of a building. In fact, models introduced in CODYRUN enable the determination of indoor daylight values at any point on a virtual surface grid representing a interior horizontal surface (including the work plane or the roof).

Experiments done on the LGI cell (fig. 13) allowed the comparison of simulated daylight values with those measured experimentally. Measurements of internal illuminance were made at five points ($A_1$, $A_2$, $A_3$, $A_4$ and $A_5$) in roofing (of height 0.01 meter above the floor). The sensors were aligned to the central axis of the building (cutting the glazing into two symmetrical parts) and perpendicular to single glazing plane (Fig. 14).

**Fig. 13. An Internal view of LGI (test room) and indoor luxmeter**          **Fig. 14. Position of the photocells (luxmeter) in working plane of LGI**

Values of global irradiance were converted to illuminance using a constant luminous efficacy model described in relation (15) and the accuracy of the prediction of illuminance was evaluated by comparing values simulated by CODYRUN with values measured in a test room (LGI cell).

Figure 15 shows two curves obtained at point $A_3$ on the ground for a given day in a dynamical behaviour of illuminance with hourly time step. The curve ('Measures.' curve) shows the measured values at point $A_3$ and the curve ('CODYRUN' curve) represents the simulated values in CODYRUN at this particular point, from global luminous efficacy model.

**Fig. 15. Simulation (CODYRUN) / Measurement comparisons for indoor daylighting in LGI cell at point $A_3$**

In this case, Mean Relative Error between simulation and measurements is 9.9 % in dynamic conditions. It is important to know that it is very difficult to characterize dynamic conditions for indoor illuminance phenomenon (see reference [32] and [33]). So, the result obtained is correct. For more information about this study, refer to Fakra's thesis [1]. The detailed study of the indoor daylighting simulation from the software CODYRUN and the associated experimental validation will be submitted soon.

## 6. Conclusion

Generally, meteorological station in the Indian Ocean does not measure solar illuminance. So it is very difficult to study visual comfort in the building or reduce energy consumption from indoor electrical lighting.

The objective of this study is to give the possibility for the software CODYRUN to be able to simulate outdoor solar illuminance starting from the knowledge of the luminous efficacy and data of solar irradiance. This would make it possible to study visual comfort and energy consumption in the building.

Two parts are presented in this paper. On one hand, this work defines three typical day-types (overcast, clear and intermediate) and the method used to classify these days. Four indices (found in the scientific bibliography and used in sky conditions) for classification have been used to determine specific day-types (overcast, intermediate and clear). To conclude SR (Sky Ratio) index can be used to characterize typical days at Saint Pierre (Reunion Island).

On the other hand, average values for global and diffuse luminous efficacy were determined. Measurements of outdoor illuminance and solar irradiance data over six months have been analysed to establish constant average global and diffuse luminous efficacy at Saint-Pierre (Reunion Island). All climatic seasons are represented in this period. The average values of global and diffuse luminous efficacy for typical day conditions are 121,5 lm/W and 139,98 lm/W respectively.

These constants were introduced into the CODYRUN software in order to simulate outdoor daylighting. A comparison between the measured values at Saint-Pierre and simulated values, on the same site, showed that the software accurately predicted illuminance values.

It is intended to develop, in the future, analytical models of luminous efficacy (global and diffuse) for Reunion Island in order to compare the simulation results with other existing models.

A similar study for the characterization of day-types conditions will be applied to other regions including Indian Ocean Island (Mauritius, Seychelles, Madagascar, Comoros, etc) using locally available database.

## Acknowledgments

This research was supported by Région Réunion. Special thanks are extended to the Région Réunion for financial support of the experimental equipment.